\documentclass[aip,jmp,amsmath,amssymb,reprint]{revtex4-1}
\usepackage{graphicx}
\usepackage{dcolumn}
\usepackage{bm}
\usepackage{color}

\begin{document}

\preprint{AIP/123-QED}

\title[Exciton-Polariton Lasing in Selenide Micropillars]{Polariton lasing in high-quality Selenide-based micropillars in the strong coupling regime}

\author{T. Klein}
\affiliation{Institute of Solid State Physics, Semiconductor Epitaxy, University of Bremen, P.O. Box 330440, 28334 Bremen, Germany}

\author{S. Klembt}
\email{sebastian.klembt@neel.cnrs.fr}
\affiliation{Institut N\'{e}el, Universit\'{e} Grenoble Alpes and CNRS, B.P. 166, 38042 Grenoble, France}

\author{E. Durupt}
\affiliation{Institut N\'{e}el, Universit\'{e} Grenoble Alpes and CNRS, B.P. 166, 38042 Grenoble, France}

\author{C. Kruse}
\affiliation{Institute of Solid State Physics, Semiconductor Epitaxy, University of Bremen, P.O. Box 330440, 28334 Bremen, Germany}

\author{D. Hommel}
\affiliation{Institute of Solid State Physics, Semiconductor Epitaxy, University of Bremen, P.O. Box 330440, 28334 Bremen, Germany}

\author{M. Richard}
\affiliation{Institut N\'{e}el, Universit\'{e} Grenoble Alpes and CNRS, B.P. 166, 38042 Grenoble, France}

\date{\today}

\begin{abstract}
We have designed and fabricated all-epitaxial ZnSe-based optical micropillars exhibiting the strong coupling regime between the excitonic transition and the confined optical cavity modes. At cryogenic temperatures, under non-resonant pulsed optical excitation, we demonstrate single transverse mode polariton lasing operation in the micropillars. Owing to the  high quality factors of these microstructures, the lasing threshold remains low even in micropillars of the smallest diameter. We show that this feature can be traced back to a sidewall roughness grain size below 3 nm, and to suppressed in-plane polariton escape.

\end{abstract}

\pacs{78.67.-n,71.36.+c,78.45.+h,78.55.Et}


\maketitle

In the last two decades exciton-polaritons in microcavities (MCs) \cite{weisbuch_observation_1992, deng_polariton_2003} have attracted considerable interest, mostly because owing to their half-light half-excitonic nature, they behave like a weakly interacting driven-dissipative bosonic quantum fluid \cite{carusotto_quantum_2013}. From an applied quantum optics point-of-view, polaritons also offer very interesting perspectives. It has been shown theoretically that by squeezing the polaritonic wavefunction into a small volume, like in micropillars, the polariton-polariton nonlinearity can become large enough to play the role of a quantum filter/emitter for light, without resorting to single emitters like atoms or quantum dots \cite{verger_polariton_2006}.
Refined etching and microstructuring techniques have been developed for GaAs-based microcavity, allowing the fabrication of high quality micropillars \cite{gutbrod_weak_1998,bajoni_polariton_2008}, mesas \cite{kaitouni_engineering_2006}, as well as advanced polaritonic circuits elements like waveguides, interferometers, optical gates \cite{wertz_spontaneous_2010,nguyen_realization_2013,sturm_all-optical_2014} and lattices with direct applications for quantum simulations \cite{kim_2011,tanese_fractal_2014,boulier_polariton-generated_2014,winkler_2015,nguyen_acoustic_2015}. This approach is likely to be successful in the upcoming years, however, for practical use, its drawback is to be stuck to cryogenic temperatures. A way around this problem is the use of large bandgap materials, where the exciton binding energy is larger, and hence stable at room temperature. But the price to pay is twofold: less mature etching techniques, and a weaker nonlinearity. Interestingly, \textit{Liew} and coworkers have shown that the quantum regime of the nonlinearity can be reached even with a small polaritonic nonlinearity by exploiting quantum interferences in a coupled micropillar pair geometry \cite{liew_single_2010}.

With this idea in mind, ZnSe or CdTe based microstructures in the strong coupling regime have a great potential: On the one hand, they have an excitonic binding energy in principle large enough to be able to maintain the strong coupling regime at room temperature \cite{saba_high-temperature_2001}. On the other hand, keeping in mind that the excitonic nonlinearity scales like the Bohr radius\cite{ciuti_role_1998} $a_B$ , the latter is still large enough in these materials to provide a sizeable nonlinearity \cite{saba_high-temperature_2001}. Finally, ZnSe has the advantage over CdTe to exhibit a polaritonic optical transition in the blue region of the spectrum ($440\,$nm versus $740\,$nm respectively), thus allowing to squeeze the polaritonic wavefunction into smaller diameter as we will see further on.

In this letter, as a first step along this route, we report on the fabrication of high quality ZnSe-based micropillars in the strong coupling regime. We find discretized transverse polariton modes, that are well resolved spectrally and in momentum space, and display a quality factor in excess of 5000 down to 2800 for the smallest diameter. Upon weak pulsed excitation, polariton lasing is achieved at cryogenic temperature in each micropillar. We find a rather low excitation threshold, which is particularly striking for the lowest diameter. We demonstrate that this feature results from the suppression of the in-plane polariton escape, which is a detrimental loss channel in planar microcavities.

\begin{figure*}[t]
\centering
\includegraphics[width=\textwidth]{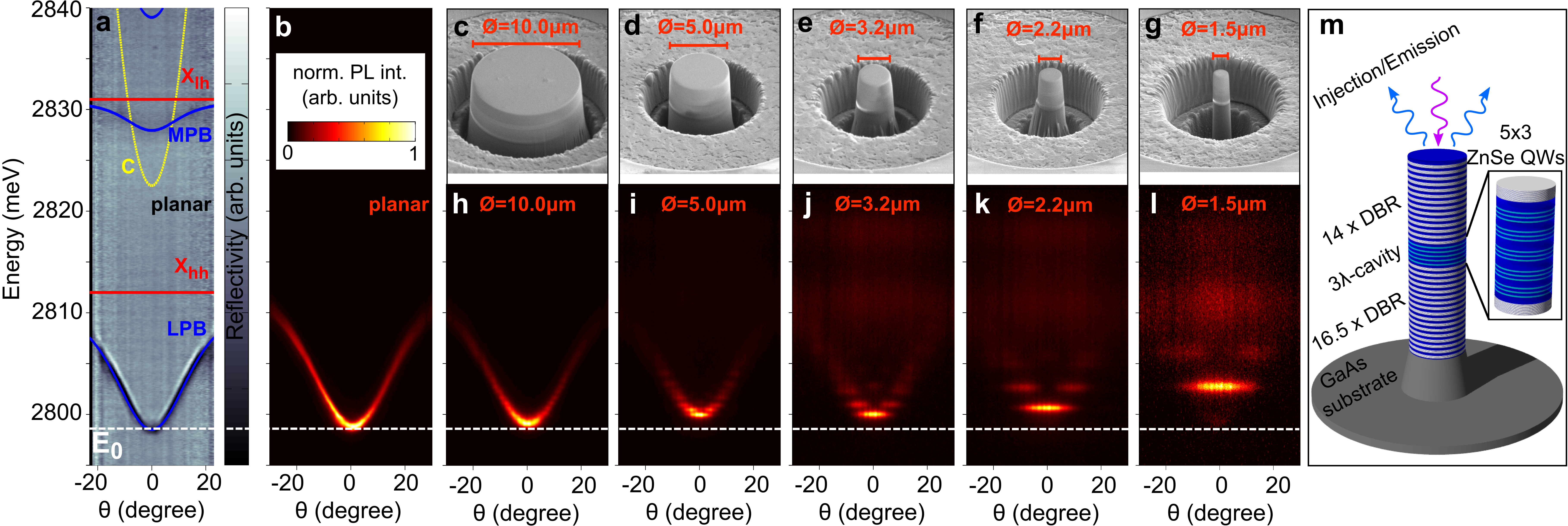}
\caption{(Color online) (a) Reflectivity as a function of energy and angle of the planar MC adjacent to the micropillars. The middle and lower polariton branches are well identified. The solid line is a three-oscillator fit of the dispersion. Calculation parameters: $\hbar\Omega_R=(32\pm 2)\,$meV, $\hbar\Omega_{lh}=(13.0\pm 1.5)\,$meV (light hole exciton Rabi splitting), $n_c=(2.15\pm 0.15)$ (cavity background index), $E_{hh}=(2812 \pm 1)\,$meV, $E_{lh}=2831\,$meV, $E_c=(2822.5 \pm 2.0)\,$meV (bare cavity energy). (b) Measured PL emission intensity below threshold as a function of energy and angle of the planar sample (linear color scale from black to white). (c)-(g) SEM images of the prepared micropillars. (h)-(l) Measured  PL emission intensity below threshold for micropillars with diameters of $10.0\,\mu$m, $5.0\,\mu$m, $3.2\,\mu$m, $2.2\,\mu$m, and $1.5\,\mu$m respectively. (m) Sketch of a ZnSe MC micropillar.}
\label{fig1}
\end{figure*}


The microcavity that we investigate in this letter is similar to that used in another work \cite{klembt_exciton-polaritons_2014}: it consists of a 16.5-fold lower distributed Bragg reflector (DBR), a 3$\lambda$-cavity with 5$\times$3 ZnSe quantum wells (QWs) placed in the antinodes of the electric field, and a 14-fold upper DBR. The high-index material of the DBRs consists of ZnMgSSe and the low-index material is a short period superlattice of MgS/ZnCdSe (further growth details available elsewhere \cite{klembt_high-reflectivity_2011, sebald_strong_2012}). Previous X-ray diffraction and transmission electron microscopy investigations show that the structure is fully lattice matched to the GaAs substrate and reveals a low defect density compared to state of the art II-VI materials \cite{klembt_structural_2013}. A set of micropillars with diameters of  \O\,=\,10\,$\mu$m, 5\,$\mu$m, 3.2\,$\mu$m, 2.2\,$\mu$m, and 1.5\,$\mu$m have been etched down through this microcavity using a focused ion beam. The micropillars design is shown in Fig.\,\ref{fig1} m, and those actually investigated are shown in Figs.\,\ref{fig1} c-g. A multistep etching protocol has been developed to minimize the etching damages on the micropillar sidewalls.

Within an angle-resolved micro-reflectivity measurement, we first characterized the planar microcavity properties. A typical angle-resolved reflectivity spectrum is shown in Fig.\,\ref{fig1} a. A three coupled oscillator model fit (involving the 1s heavy-hole excitons and the 1s light hole excitons \cite{astakhov_charged_1999}) yields a heavy hole exciton Rabi splitting $\hbar\Omega_R=(32\pm 2)\,$meV. In the region of the micropillars, a detuning $\delta=E_C-E_X=+(10.5\pm 3.0)\,$meV is found.

\begin{figure}[hbt]
\includegraphics[width=0.44\textwidth]{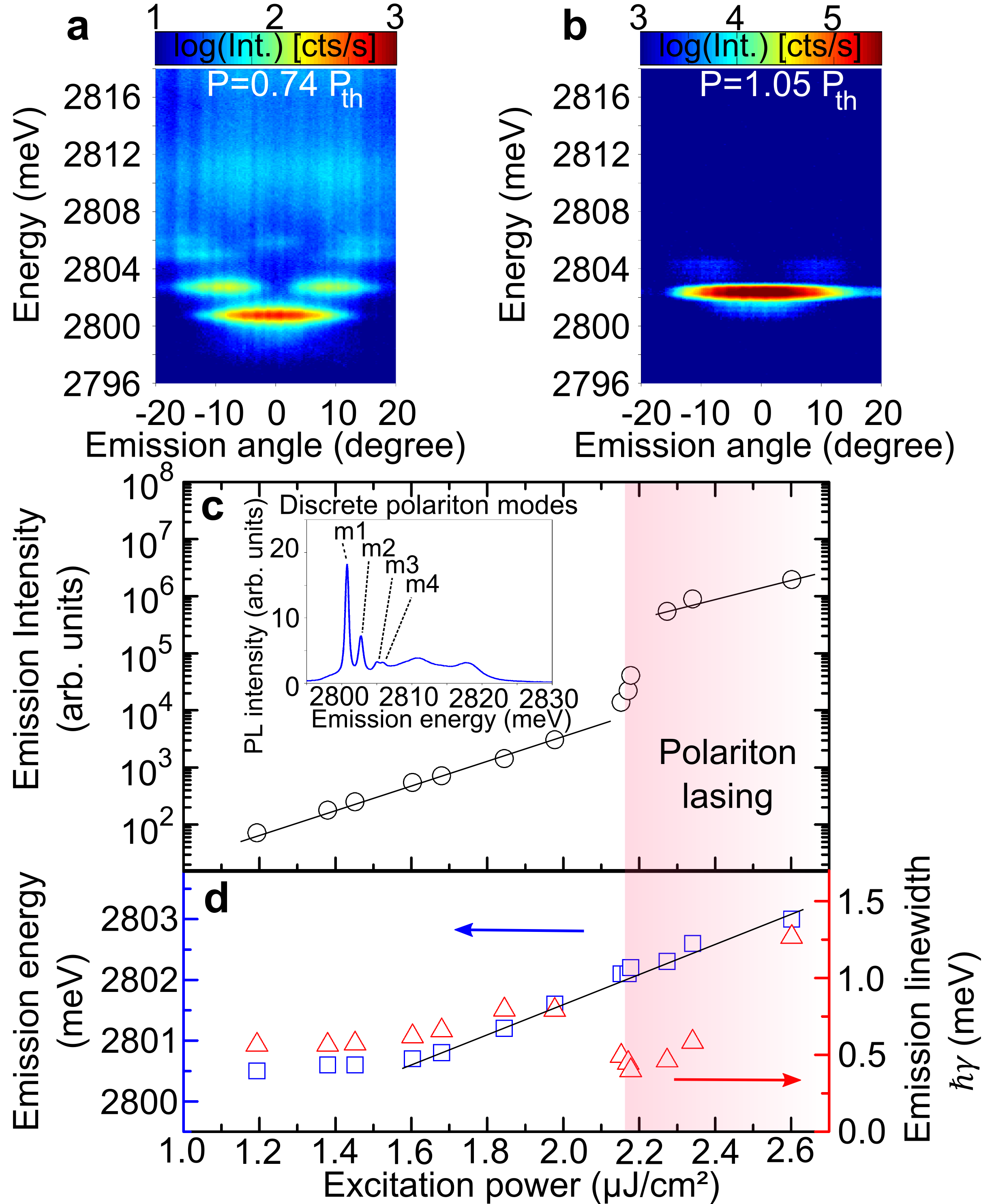}
\caption{(Color online) Measured polariton PL intensity (log. color scale) of the 2.2\,$\mu$m micropillar as a function of energy and emission angle below (a) and above (b) the polariton lasing threshold (P$_{\text{thr}}=2.16\,\mu$J/cm$^2$). Integrated PL intensity (c), and emission energy and linewidth (d) measured on the lowest energy emission mode $m1$ of the 2.2\,$\mu$m-micropillar as a function of the excitation power. The pink region on the plots right side materializes the power range $P$ above the lasing threshold $P_{\text{thr}}$=2.16\,$\mu$J/cm$^2$. Inset: angle-integrated PL spectrum measured on a 2.2\,$\mu$m-micropillar with an excitation power of  $P$=1.60\,$\mu$J/cm$^2$.}
\label{fig2}
\end{figure}

We then realized microphotoluminescence (PL) measurements of each micropillar, using a frequency-doubled picosecond-pulsed Ti-Sapphire laser, focused by a microscope objective into a Gausssian spot approximately 12\,$\mu$m in diameter. Excitation is carried out non-resonantly through the first high-energy mode of the Bragg mirrors at $E_{Laser} \approx 2950$\,meV. The sample was kept in a Helium-flow cryostat at a constant temperature of $T\,=\,6$\,K.

The angle-resolved PL spectra of the micropillars in the weak excitation regime are shown in Fig.\,\ref{fig1} h-l, alongside that of the nearby planar microcavity (Fig.\,\ref{fig1} b). Owing to the narrow polaritons linewidth, transverse state discretization is well resolved up to a diameter of $10\,\mu$m. We also observe the expected blueshift of the polariton ground state (that centered around $k_\parallel=0$) for increasing lateral confinement of the polaritonic mode within the micropillars \cite{kaitouni_engineering_2006}. This blueshift reaches $4.2\,$meV for the (smallest) micropillar diameter of $1.5\,\mu$m. The strong coupling regime is thus unambiguously conserved in these micropillars since the bare confined cavity mode is more than $25\,$meV above the polariton ground state in the planar microcavity. Qualitatively, the confined polaritons exhibit the expected angular emission pattern as predicted by \textit{Kaitouni} and coworkers \cite{kaitouni_engineering_2006}. This points towards the fact that in spite of the etching step, the micropillars exhibit a low sidewall roughness.

\begin{figure}[bt]
\centering
\includegraphics[width=0.43\textwidth]{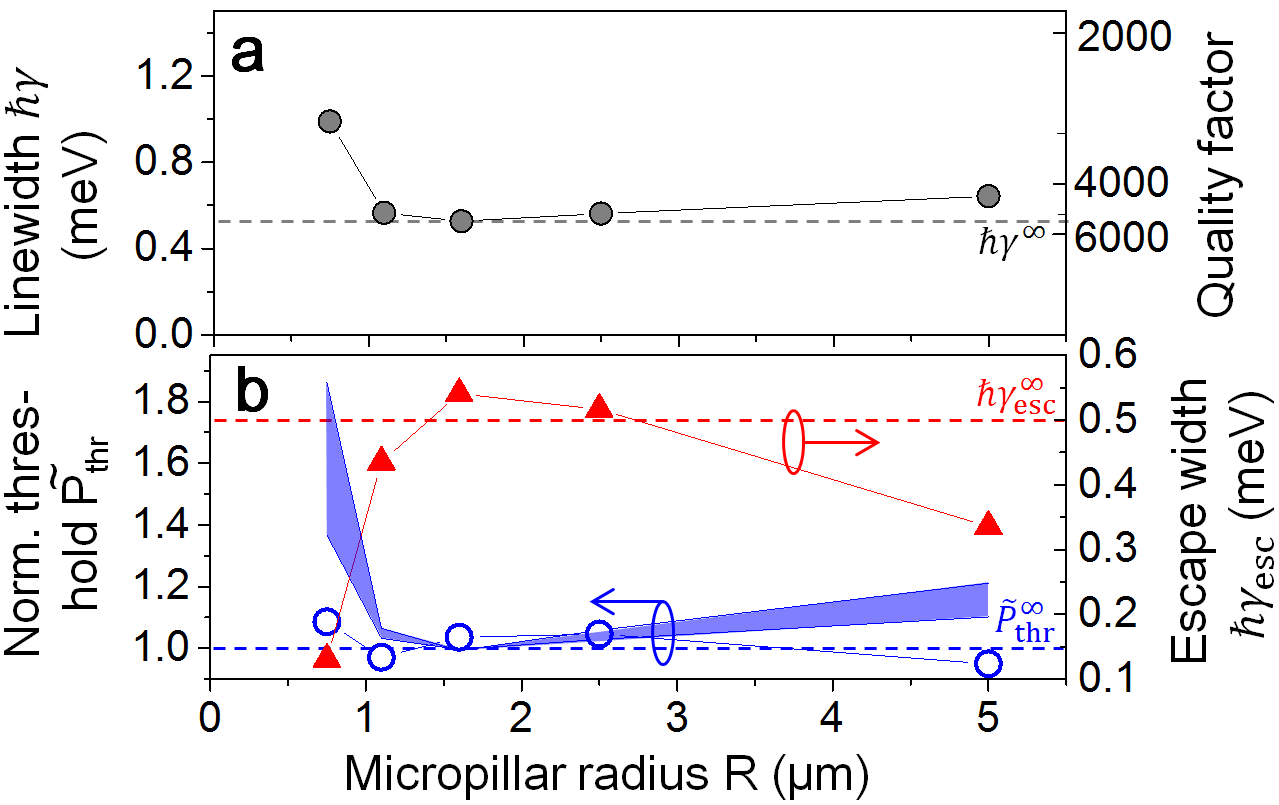}
\caption{(a) Linewidth $\hbar\gamma$ (in PL) of the polariton ground-state versus micropillar radius $R$. (b) Left axis: Measured (open blue symbols) and calculated (blue surface, $0\leq A_0\leq 0.5$) excitation threshold, for a calculation accounting only for the radiative losses $\gamma$. On the right axis, the derived polariton losses by in-plane escape $\hbar\gamma_\text{esc}$ are plotted. The horizontal dashed lines show the measured values in the planar microcavity.}
\label{fig3}
\end{figure}

An analysis of the linewidth versus diameter, extracted from the PL data, and shown in Fig.\,\ref{fig3} a, can yield a quantitative estimation of the latter. \textit{Borselli} and coworkers have shown that \cite{borselli_beyond_2005}
\begin{equation}
l_r^6=\frac{3\lambda_0^3 R^2 h}{8\pi^{5/2}n_p^2},
\end{equation}
where $\lambda_0=443\,$nm is the polariton wavelength in vacuum, $R$ and $h=1.4\,\mu$m are the micropillar radius and the mode effective height respectively, and $n_p=2\times2.15$ is the polaritonic effective index. This estimation results in a characteristic roughness size as low as $l_r=2.3\pm0.7$nm. This very small grain size is comparable to well established inductively coupled plasma and focused ion beam etching techniques in Si and III-V semiconductors, where grain sizes between 1.5 to 4.0\,nm have been reported \cite{borselli_beyond_2005,bae_characterization_2003,song_smoothing_2007}. This result shows that our FIB etching technique meets state-of-the-art standards.

Figs.\,\ref{fig2} a and b show the angle-resolved PL spectra of a 2.2\,$\mu$m diameter micropillar below and above the polariton lasing threshold, respectively. Despite the very high population contrast, we can still see the first polariton excited state above threshold, with an unchanged momentum-space pattern as compared to below threshold. A two-body interaction induced blueshift of $1.6\,$meV is observed at threshold, i.e. an order of magnitude lower than that expected in the case where the strong coupling regime would break up.



A detailed analysis of the polariton ground-state PL (labelled $m1$ in the inset) in this small diameter micropillar across the lasing threshold is shown in Fig.\,\ref{fig2} c-d. Fig.\,\ref{fig2} c shows that at a threshold power $P_{\text{thr}}=2.16\,\mu$J/cm$^2$ (as measured at the entrance of the microscope objective), a two orders of magnitude increase of the emission from the ground state is observed within $7\%$ of excitation power increase. Moreover, as shown in Fig.\,\ref{fig2} d, the linewidth decreases from $\Delta E_{m1}\,=\,0.56\,$meV (full width at half-maximum) at low excitation power to $\Delta E_{m1}^{\text{thr}}\,=\,0.40\,$meV at threshold, and then increases again at higher power due to two-body scattering decoherence. This is the typical signature of polariton stimulated relaxation. The limited linewidth narrowing is the result of pulsed excitation and time-integrated detection. The measured blueshift versus excitation power is shown in Fig.\,\ref{fig2} d. It increases linearly below (starting from $P\sim\,0.74\,P_{\text{thr}}$), at, and above the threshold without notable slope change over the whole range. This is the expected behaviour when the latter is due to two-body scattering, involving excitons below threshold, and excitons and polaritons above threshold. This is unlike the weak coupling regime, where the electron-hole reservoir density gets clamped by the stimulation at and above threshold, resulting in a sharp slope change at threshold, i.e. a saturation of the blueshift upon increasing power further above threshold \cite{bajoni_polariton_2008}. These features match what has been reported already in GaAs micropillars by \textit{Bajoni et al.} \cite{bajoni_polariton_2008} and clearly show that the micropillars are lasing in the strong coupling regime.

Interestingly, $P_{\text{thr}}$ is rather low for a non-GaAs polaritonic microstructure. Indeed, in similar experimental conditions (non-resonant, pico- or femto-second pulse excitation, $6\,\text{K}<T<10\,\text{K}$), a typical excitation threshold of $4\,\mu\text{J/cm}^{2}\leq P_\text{thr}\leq 30\,\mu\text{J/cm}^{2}$ can be found in the literature in GaAs-based planar microcavities \cite{deng_polariton_2003,tempel_characterization_2012,tempel_temperature_2012}, of $62\,\mu$J/cm$^{2}$ in CdTe planar microcavities \cite{richard_experimental_2005}, and of $8\,\mu$J/cm$^{2}$ in high quality ZnO microwires \cite{trichet_2013}.
Fig.\,\ref{fig3} b shows how the excitation threshold changes when passing from lasing in a planar area of the microcavity to lasing in micropillars of decreasing diameters and a similar detuning (within the small optical confinement energy). We use a simple three-level rate equation model to describe polariton lasing in the general case :
\begin{equation}
\begin{cases}
\partial N_r/\partial t = P-\gamma_{nr}N_r-\gamma_r(N_p+1)N_r^2\\
\partial N_p/\partial t = \gamma_r(N_p+1)N_r^2-\gamma N_p,
\end{cases}
\end{equation}
where $N_r$ and $N_p$ are the reservoir (excitons) and ground-state polariton population respectively; $\gamma_{nr}$, $\gamma_r$ and $\gamma$ are the reservoir non-radiative loss rate, the exciton to polariton relaxation rate, and the polariton radiative rate respectively. $P$ is the external excitation rate. A relaxation term quadratic in the reservoir population $N_r$ is taken in agreement with our measurements. According to this model, within the steady-state regime approximation, the excitation threshold versus diameter depends mostly on $\gamma$ as
\begin{equation}
\tilde{P}_{\text{thr}}(R)=A_0\tilde{\gamma}(R)+ \tilde{\gamma}(R)^{1/2}(1-A_0),
\end{equation}
where $\tilde{P}_{\text{thr}}(R)=P_{\text{thr}}(R)/P_{\text{thr}}^\infty$ is the threshold excitation power in a micropillar of radius $R$ normalized to $P_{\text{thr}}^\infty$, the threshold in the planar microcavity; $\tilde{\gamma}(R)=\gamma(R)/\gamma^\infty$ is the normalized polariton linewidth, and $A_0=\gamma^\infty/P_{\text{thr}}^\infty$, where $P_{\text{thr}}^\infty$ is the excitation rate at threshold. Surprisingly, as shown in Fig.\,\ref{fig3} b, for any realistic values of $A_0 \ll 1$ (blue area in the plot), and using the measured $\gamma(R)$ reported in Fig.\,\ref{fig3} a, the model significantly overestimates $\tilde{P}_{\text{thr}}(R)$. It looks as if the micropillars would have \emph{lower} losses than the non-etched planar area of the microcavity. The above model assumes a radius-independent non-radiative loss rate $\gamma_{nr}$ of the reservoir; we thus considered making it radius-dependent, however, this attempt yields a nonphysical result, namely that the non-radiative recombination rate would decrease for decreasing diameter.

In order to explain our measurement, we need to invoke a polariton loss channel that does not contribute to its linewidth in PL. It was pointed out already that confinement is enhancing polariton relaxation by phonons\cite{paraiso_2009}. However, in the polariton lasing regime, it is unlikely the dominant mechanism of stimulated relaxation, rather driven as discussed above by two-body scattering. Another mechanism, better known in the context of polariton lasing in planar microcavities, is more likely: unlike for regular lasers, the gain medium is always transparent for polaritons, even in the non-excited regions. Therefore, under spatially inhomogeneous excitation, polaritons generated at the center of the excitation spot experience a repulsive potential gradient (due to ta stronger blueshift in the center) that pushes them away from the laser spot center \cite{richard_experimental_2005,wouters_2008,wertz_spontaneous_2010}. The polaritonic field effectively driving the stimulated relaxation is thus lowered by this mechanism. In micropillars, this loss channel is increasingly suppressed for decreasing diameter since non-zero momentum states (i.e. excited states) are split away from the ground state by an increasing energy gap. Thus, for sufficiently high optical quality factors like reported in this work, this effect competes with the radius dependent linewidth, to fix the polariton laser threshold.

We verified this assumption by adding a polariton escape rate $\gamma_\text{esc}(R)$ to the model. By fitting the measured $\tilde{P}_{\text{thr}}(R)$ and leaving this parameter free we could come up with two interesting features: (i) the polariton escape rate $\gamma_{\text{esc}}(R)$ versus diameter shown in Fig.\,\ref{fig3} b (right axis), and (ii) the minimum escape rate $\hbar\gamma_\text{esc}^\infty=0.42$meV in the planar microcavity, which is needed in order to be consistent with the measurements. As expected, in micropillars, the thus derived escape rate is markedly lower than in the planar case, in particular for the smallest diameter. Note that the observed non-monotonous behaviour of $\gamma_{\text{esc}}(R)$ is most likely a consequence of the local disorder in each micropillar, which also contributes to polariton in-plane redistribution.

In conclusion, we have shown that we could fabricate high quality Selenide based micropillars in the strong coupling regime, where low threshold polariton lasing at cryogenic temperature is achieved. We could show that when the polariton optical loss rate is low enough, the polariton lasing mechanism in micropillars benefits from the suppression of in-plane polariton escape. A recent work of ours \cite{klembt_exciton-polaritons_2014} as well as some preliminary experiments suggest that the strong coupling regime should be conserved, and polariton lasing still functional, up to room temperature. Such a device would be a serious progress towards the actual use of polaritons in applied quantum optics.

 The authors acknowledge support by the ERC StG contract Nb 258608. Stimulating discussions with J. Claudon and L. S. Dang, as well as technical support by C. Bouchard and L. Del-Rey are warmly acknowledged.



\end{document}